\documentclass[twocolumn,showpacs,oneside,amsmath,amssymb,prl]{revtex4}
\usepackage{amsmath}
\usepackage{graphicx}
\usepackage{dcolumn}
\usepackage{bm}

\begin{document}

\title{Quantum Phase Transition in Hall Conductivity on an Anisotropic Kagom\'{e} Lattice}

\author{Shun-Li Yu}
\author{Jian-Xin Li}
\email{jxli@nju.edu.cn}
\author{Li Sheng}
\affiliation{National Laboratory of Solid State
Microstructures and Department of Physics, Nanjing University,
Nanjing 210093, China}

\date{\today}

\begin{abstract}
We study the quantum Hall effect(QHE) on the Kagom\'{e} lattice with
anisotropy in one of the hopping integrals. We find a new type of
QHE characterized by the quantization rules for Hall conductivity
$\sigma_{xy}=2ne^{2}/h$ and Landau Levels $E(n)=\pm
v_{F}\sqrt{(n+1/2)\hbar Be}$ ($n$ is an integer), which is different
from any known type. This phase evolves from the QHE phase with
$\sigma_{xy}=4(n+1/2)e^{2}/h$ and $E(n)=\pm v_{F}\sqrt{2n\hbar Be}$
in the isotropic case, which is realized in a system with massless
Dirac fermions (such as in graphene). The phase transition does not
occur simultaneously in all Hall plateaus as usual but in sequence
from low to high energies, with the increase of hopping anisotropy.
\end{abstract}


\maketitle

The quantum Hall effect (QHE) is a remarkable transport phenomena in
condensed matter physics~\cite{Prange}. There are three kinds of
integer QHE in the known materials. One is the conventional
integer QHE occurring in two-dimensional (2D) semiconductor
systems, where the successive filling of the Landau levels leads to an
equidistant ladder of quantum Hall plateaus at integer filling
$n=0,\pm1,\pm2,\cdots$, with a quantized value
$\sigma_{xy}=2ne^{2}/h$~\cite{Prange}. The second is the
unconventional QHE observed in graphene, where charge carriers
mimic the massless Dirac fermions, so that the Hall conductivity
is half-integer quantized $\sigma_{xy}=4(n+1/2)e^{2}/h$ due to
a Berry phase shift $\pi$ at the Dirac points
~\cite{Novoselov1,Zhang,Haldane,Zheng,Gusynin,Sheng,Neto}. The
third occurs in bilayer graphene, where the charge carriers have a
parabolic energy spectrum but are chiral with a Berry's phase
$2\pi$. Therefore, the Hall conductivity follows the same ladder as
in conventional 2D electron gases, but the plateau at zero level
is absent~\cite{Novoselov2,McCann}.

In this Letter, we demonstrate a new kind of integer QHE on the
anisotropic Kagom$\acute{e}$ lattice and the quantum phase
transition relating it to the unconventional QHE for the massless
Dirac fermions. The Kagom$\acute{e}$ lattice has recently attracted
considerable interest due to its higher degree of frustration. It is
the line graph of the honeycomb structure in view of the graph
theory~\cite{Mielke}.  The three-band electronic structure [Fig.1]
is composed of one flat band and two dispersive bands. The latter
has the same form as that in graphene~\cite{Haldane}, and the two
bands touch at two inequivalent Dirac points forming massless Dirac
fermions. As a result, an unconventional QHE with
$\sigma_{xy}=4(n+1/2)e^{2}/h$ is realized on the isotropic
Kagom$\acute{e}$ lattice. Assuming one of the three hopping
integrals, which is denoted by $t_{23}$, can take a different value
from the two others, we find a quantum phase transition for the Hall
conductivity from the unconventional form
$\sigma_{xy}=4(n+1/2)e^{2}/h$ to $\sigma_{xy}=2ne^{2}/h$. Though the
latter phase shows the same Hall conductivity as that found in
conventional 2D semiconductors, it has the following non-trivial
properties. i) The phase is characterized by a new quantization rule
of Landau levels (LL) with $E(n)=v_{F}\sqrt{(n+1/2)\hbar Be}$, in
contrast to $E(n)=(n+1/2)\hbar\omega_{c}$ with $\omega_{c}=eB/m$ for
the free-fermion QHE systems, $E(n)=v_{F}\sqrt{2n\hbar Be}$ for the
single-layer graphene and $E(n)=\sqrt{n(n-1)}\hbar\omega_{c}$ for
the bilayer graphene. ii) The quantum phase transition does not
occur simultaneously in all Hall plateaus as usual but in sequence
from low to high energies with the increase of anisotropy. iii) The
quantum phase transition occurs only in the case of $t_{23}<t_{12}$
($t_{12}=t_{13}$). In the other case of $t_{23}>t_{12}$, the
unconventional QHE realized in the isotropic system remains at least
for the largest anisotropy we considered here, namely $t_{23}=2
t_{12}$. This kind of quantum phase structures controlled by the
anisotropy of the hopping parameters is also in stark contrast to
that in the honeycomb lattice(graphene), where the unconventional
QHE evolves into the conventional one for the strong $t_{23}$
($t_{23}>t_{12}$) regime, while no phase transition occurs for the
weak $t_{23}$ ($t_{23}<t_{12}$)
regime~\cite{Sato,Hasegawa,Dietl,Kohmoto}. Therefore, the quantum
phase transition demonstrated here has no known analogues and
presents an intriguing case for experimental studies on QHE.

\begin{figure}
  \includegraphics[scale=0.38]{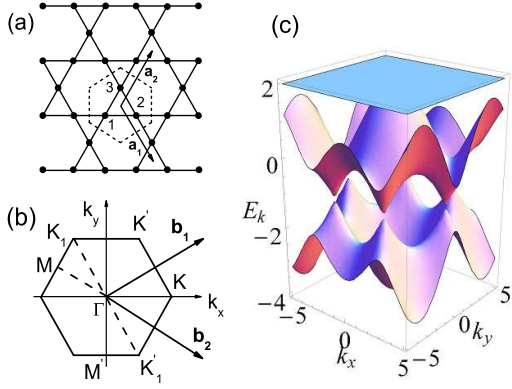}\\
  \caption{(color online) (a) Lattice structure of the Kagom$\acute{e}$ lattice. The region
  enclosed by the dash lines represents the Wigner-Seitz unit cell. $\mathbf{a}_{1}$ and
$\mathbf{a}_{2}$ are the lattice
  unit vectors. The lattice sites in each unit cell are labeled by 1, 2, and 3. (b) The first
  Brillouin zone. $\mathbf{b}_{1}$ and $\mathbf{b}_{2}$ are the reciprocal-lattice vectors.
  (c) The electronic dispersion for $t_{12}=t_{23}=t_{31}=1.0$. The Dirac cones are located
  at $K$($K_{1}$) and $K'$($K'_{1}$) points as labeled in Fig.(b).}
  \label{fig1}
\end{figure}

We start from the tight-binding model on a 2D metallic
Kagom$\acute{e}$ lattice,
\begin{eqnarray}
\hat{H}=-\sum_{\langle
ij\rangle,\sigma}(t_{ij}\hat{c}^{\dag}_{i\sigma}\hat{c}_{j\sigma}+\mathrm{H}.\mathrm{C}.),
\label{h}
\end{eqnarray}
where $\hat{c}_{i\sigma}$($\hat{c}^{\dag}_{i\sigma}$)
annihilates(creates) an electron with spin
$\sigma$($\sigma=\uparrow,\downarrow$) on site $i$ and $t_{ij}$ is
the hopping integral between the nearest neighbors. Considering that
there are three sites in each unit cell [see Fig.~\ref{fig1}(a)], we
can write Eq.(~\ref{h}) in the momentum space as $
\hat{H}=\sum_{\mathbf{k}\sigma}\hat{\Psi}^{\dag}_{\mathbf{k}\sigma}
M(\mathbf{k})\hat{\Psi}_{\mathbf{k}\sigma}$. Where
$\hat{\Psi}_{\sigma}(\mathbf{k})=(\hat{c}_{\mathbf{k}1\sigma},\hat{c}_{\mathbf{k}2\sigma},\hat{c}_{\mathbf{k}3\sigma})$,
and $M(\mathbf{k})$ is a $3\times3$ matrix
\begin{eqnarray}
M(\mathbf{k})=\left(\begin{array}{ccc}
        0 & \varepsilon_{12}(\mathbf{k}) & \varepsilon_{13}(\mathbf{k}) \\
        \varepsilon_{21}(\mathbf{k}) & 0 & \varepsilon_{23}(\mathbf{k}) \\
        \varepsilon_{31}(\mathbf{k}) & \varepsilon_{32}(\mathbf{k}) & 0
  \end{array}\right),
\end{eqnarray}
with
$\varepsilon_{12}(\mathbf{k})=\varepsilon_{21}(\mathbf{k})=-2t_{12}\cos(\mathbf{k}\cdot\mathbf{\delta}_{1})$,
$\varepsilon_{13}(\mathbf{k})=\varepsilon_{31}(\mathbf{k})=-2t_{31}\cos(\mathbf{k}\cdot\mathbf{\delta}_{3})$
and $\varepsilon_{23}(\mathbf{k})=\varepsilon_{32}(\mathbf{k})=
-2t_{23}\cos(\mathbf{k}\cdot\mathbf{\delta}_{2})$. $\mathbf{
\delta}_{1}$, $\mathbf{\delta}_{2}$ and $\mathbf{\delta}_{3}$ are
the nearest-neighbor vectors, $\mathbf{\delta}_{1}=(1/2)\hat{x}$,
$\mathbf{\delta}_{2}=(1/4)\hat{x} +(\sqrt{3}/4)\hat{y}$,
$\mathbf{\delta}_{3}=-(1/4)\hat{x} +(\sqrt{3}/4)\hat{y}$. For the
isotropic case ($t_{12}=t_{23}=t_{31}=t$), the energy bands are,
\begin{equation}
E_{\pm}(\mathbf{k})=-t\pm t\sqrt{1+f(\mathbf{k})}, \  \
E_{3}(\mathbf{k})=2t, \label{df}
\end{equation}
where $f(\mathbf{k})=8\cos(k_{x}/2)\cos(k_{x}/4+\sqrt{3}k_{y}/4)
\cos(k_{x}/4-\sqrt{3}k_{y}/4)$, and the energy band structure is
shown in Fig.~\ref{fig1}(c). The two dispersive bands $E^{+}$ and
$E^{-}$ contact at the Dirac point $K=(4\pi/3,0)$[or
$K'=(2\pi/3,2\pi/\sqrt{3})$]. Near the Dirac point
$\mathbf{q}=\mathbf{k}-\mathbf{K}$, the electrons behave as Dirac
fermions with the approximated dispersion,
\begin{eqnarray}
E_{\pm}=-t\pm v_{F}|\mathbf{q}|+O[(q/K)^{2}],
\end{eqnarray}
$v_{F}=3\sqrt{2}t/4$ is the Fermi velocity. Thus, except for an
additional flat band $E_{3}$, the two dispersive energy bands are
similar to those in graphene. As a result, an unconventional QHE
as that found in graphene is expected on the isotropic
Kagom$\acute{e}$ lattice around the Fermi level $\mu=-t$
(corresponding to the electron filling $1/3$).

In an uniform magnetic field applied perpendicular to the sample
plane, the tight-binding Hamiltonian is,
\begin{eqnarray}
\hat{H}=-\sum_{\langle
ij\rangle,\sigma}(t_{ij}e^{i\alpha_{ij}}\hat{c}^{\dag}_{i\sigma}\hat{c}_{j\sigma}+\mathrm{H}.\mathrm{C}.).
\label{mh}
\end{eqnarray}
The Hamiltonian Eq.(5) will be diagonalized numerically on a finite
lattice with size $N=3\times L_{1}\times L_{2}$(factor three counts
the three sites in each unit cell). The magnetic flux per triangular
is chosen to be $\phi=\sum_{\triangle}\alpha_{ij}=\frac{2\pi}{M}$,
with $M$ an integer, then the total flux $\Phi=\frac{16\pi N}{3M}$
through the lattice is taken to be an integer to satisfy the
periodic boundary condition. Typically, $N=3\times320\times320$ and
$\phi=\frac{2\pi}{2560}$ are used in numerical calculations. After
the diagonalization, the Hall conductivity $\sigma_{xy}$ is
calculated with the Kubo formula
\begin{equation}
\sigma_{xy}=A\sum_{\varepsilon_{\alpha}<E_{F}}\sum_{
\varepsilon_{\beta}>E_{F}}
\frac{\langle\alpha|v_{x}|\beta\rangle\langle\beta|v_{y}|\alpha\rangle-\langle\alpha|v_{y}|\beta\rangle
\langle\beta|v_{x}|\alpha\rangle}{(\varepsilon_{\alpha}-\varepsilon_{\beta})^{2}},
\end{equation}
where $A=i2e^{2}/S\hbar$ with $S$ the area of the system, $v$ is the
velocity, $\varepsilon_{\alpha}$ and $\varepsilon_{\beta}$ are the
corresponding eigenvalues of the eigenstates $|\alpha\rangle$ and
$|\beta\rangle$. In the following, the hopping integrals
$t_{12}=t_{31}=1.0$ are used as the energy unit and $t_{23}$ to
acount for the anisotropy.

\begin{figure}
  \includegraphics[scale=0.38]{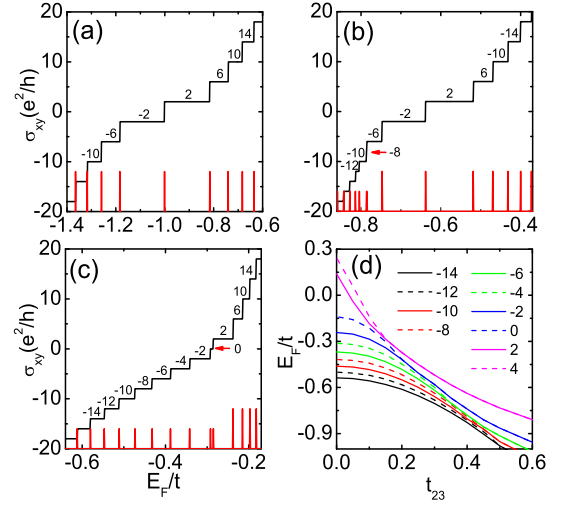}\\
  \caption{(color online) (a)-(c) Hall conductivities for different $t_{23}$ as a function of the
  Fermi energy and the corresponding density of states which is denoted by red vertical lines [(a) $t_{23}=1.0$.
   (b) $t_{23}=0.4$. (c) $t_{23}=0.15$]. The
  numbers on each plateau denote $\nu$ in $\sigma_{xy}=\nu e^{2}/h$.
  The red arrows in (b) and (c) indicate the transition point from $g_{s}=4$ to
$g_{s}=2$. (d) Phase diagram showing the gradual splitting of the
Hall plateaus with the decrease of $t_{23}$. The solid lines denote
the original plateaus and the dashed lines the new plateaus induced
by the anisotropy.}
  \label{fig2}
\end{figure}
In Fig.~\ref{fig2}, the Hall conductivity
$\sigma_{xy}=\nu\frac{e^{2}}{h}$ and the electron density of
states(DOS) are plotted for the different $t_{23}$. For the
isotropic case with $t_{23}=1.0$, the Hall plateaus satisfy the
unconventional quantization rule $\nu=(n+1/2)g_{s}$ with a
degeneracy factor $g_{s}=4$ for each Landau level(LL) to count two
spin components and two Dirac points. The $1/2$ shift of the Hall
plateaus is due to a nonzero Berry phase at the Dirac points~\cite{Mikitik},
or can be simply explained as arising from
the existence of a zero mode as seen from the DOS shown in Fig.2(a).
This numerical result is consistent with the above analytical
calculation, showing that the QHE on the isotropic Kagom$\acute{e}$
lattice exhibits the same behavior as that in graphene.

\begin{figure}
  \includegraphics[scale=0.38]{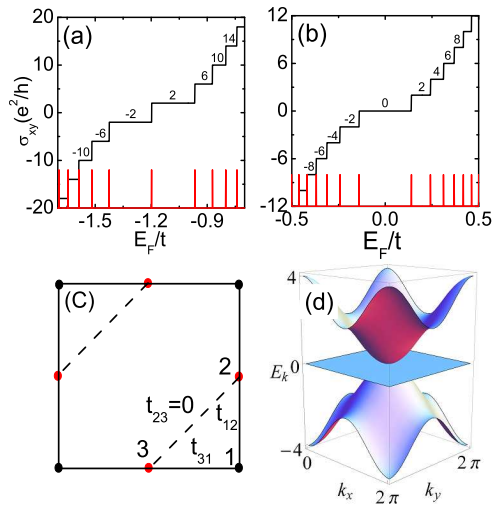}\\
  \caption{(color online) (a) and (b) are the Hall conductivity together with the DOS for
  $t_{23}=1$ and $t_{23}=0$, respectively. (c) The square lattice
  which is equivalent topologically to the Kagom$\acute{e}$
  lattice for $t_{23}=0$. (d) The energy bands for the lattice in (c).
  }
  \label{fig3}
\end{figure}

Decreasing $t_{23}$ to introduce the anisotropy in the hopping
integral, one will find that the steps in Hall conductivity split at
the mid-point gradually, as shown in  Fig.2(b) and Fig.2(c).
Concomitant with the splitting, a new Hall plateau emerges between
every two Hall plateaus and the degeneracy factor $g_{s}$ changes
from $4$ to $2$. Basically, one may expect that, upon the
introduction of the anisotropy, the rotational symmetry of the
isotropic Kagom$\acute{e}$ lattice will be broken, and consequently
the degenerate energy levels will be separated. Indeed, at the
energy levels where the step splitting in Hall conductivity occurs,
the peak of DOS (denoted as red vertical lines in Fig.2) is split
into two adjacent peaks with half a previous height. However, two
nontrivial characters exhibit here. One is that the splitting does not
happen as a whole simultaneously, but gradually from low energies to
high energies with the decrease of $t_{23}$ (corresponding to the
enhancement of the anisotropy). Moreover, whenever a splitting
occurs, a new Hall plateau emerges. But, the next splitting does not
happen in succession with the further decrease of $t_{23}$, instead
the emerging plateau will grow firstly until it satisfies the new
quantization rules for the Landau Level which will be addressed in
the following. This process can also be clearly seen from the peak
splitting in DOS. In this way, the QHE on the anisotropic
Kagom$\acute{e}$ lattice exhibits a sequent quantum phase transition
from low to high energies, as shown in the phase diagram presented
in Fig.2(d). Secondly, in the case of $t_{23}>1$ (see
Fig.~\ref{fig3}(a)), the peaks in DOS does not split at any energy
level in the reasonable parameter regime. As a result, at least for
the largest anisotropy $t_{23}=2$ considered here, the QHE shows the
same behavior to the isotropic case.

The quantum phase transition demonstrated above is in stark contrast
to that on the honeycomb
lattice(graphene)~\cite{Sato,Hasegawa,Dietl,Kohmoto}, where the
unconventional QHE changes into the conventional QHE with
$\sigma_{xy}=2ne^{2}/h$ in the strong $t_{23}$ ($t_{23}>1$) regime,
while no phase transition occurs for the weak $t_{23}$ ($t_{23}<1$)
regime. In addition, the phase transition on the anisotropic
honeycomb lattice occurs symmetrically starting from both the low
and high energies, and gradually approaches to the zero-energy
level, so that it exhibits a particle-hole symmetry.

Next, let us study the character manifested by the new phase
emerging in the quantum phase transition. This is accessible easily
in the limit $t_{23}=0$, where the Kagom$\acute{e}$ lattice is
topologically equivalent to the lattice in Fig.~\ref{fig3}(c) with
only the nearest neighbor hoppings. The three energy bands on this
lattice can be found as,
$E_{\pm}=\pm2t\sqrt{\cos^{2}(k_{x}/2)+\cos^{2}(k_{y}/2)}$ and
$E_{3}=0$, as shown in Fig.~\ref{fig3}(d). Around
 $\mathbf{K}_{0}=(\pi,\pi)$, the bands $E_{\pm}$ have the linear
dispersion $E_{\pm}(\mathbf{q})=\pm v_{F}|\mathbf{q}|$ with
$v_{F}=\sqrt{2}t$. Because the flat band $E_{3}$ crosses the
$\mathbf{K}_{0}$ point, the electrons do not behave as massless
Dirac fermions. In this respect, the Hall conductivity shows a
conventional behavior $\sigma_{xy}=2ne^{2}/h$. However, the energy
of the LL follows a new quantization rule(Fig.3(b)). This can be
obtained analytically by solving the low-energy Hamiltonian of the
system, as given by
\begin{eqnarray}
\hat{H}=\frac{v_{F}}{\sqrt{2}}\left(\begin{array}{ccc}
        0 & \hat{p}_{x} & \hat{p}_{y} \\
        \hat{p}_{x}^{\dag} & 0 & 0 \\
        \hat{p}_{y}^{\dag} & 0 & 0
  \end{array}\right),
\end{eqnarray}
under a uniform magnetic field $\mathbf{B}$ by replacing the
momentum operator $\hat{\mathbf{p}}$ by
$\hat{\mathbf{p}}+e\mathbf{A}$ with $\mathbf{A}=B(-y,0)$. The result
turns out to be,
\begin{eqnarray}
E(n)=\pm v_{F}\sqrt{(n+1/2)\hbar Be},\ n=1,2,3,\cdots.
\end{eqnarray}
This kind of distribution of the LL leads to a corresponding
distribution of the Hall plateaus, which is different from that in
conventional semiconductors $(n+1/2)\hbar\omega_{c}$, graphenes
$v_{F}\sqrt{2n\hbar Be}$ and bilayer
graphenes$\sqrt{n(n-1)}\hbar\omega_{c}$.

To understand the property of the quantum phase transition
demonstrated above, we show in Fig.4 the evolution of the energy
band with the hopping integral $t_{23}$ along the high symmetrical
directions. First, we point out that the unconventional QHE is
found numerically to be limited to a finite energy range (not
shown here), which is the region from $A$ to $A'$ (points of the
van Hove singularity) in the dispersions shown in Fig.4. Outside
this region, the QHE will exhibit a conventional behavior in Hall
conductivity. For $t_{23}<1$ [Fig.4(a)], the two Dirac points,
which are at $K$ and $K'$ points in the isotropic case, approach
each other along the $K-K'$ direction with the decrease of
$t_{23}$. Interestingly, in this process, the energy band around
the $A$ point is suppressed and lifted upwards gradually, while
that around the $A'$ point is not changed. As a result, the low
energy part will be excluded out of the unconventional regime.
Therefore, the conventional Hall conductivity emerges at these
energy levels.

On the other hand, for $t_{23}>1$ [Fig.4(b)] the two Dirac points
approach each other along the $K_{1}-\Gamma-K'_{1}$ direction [see
Fig.1(b) for illustration]. Different from the case of $t_{23}<1$,
the energy band around $A$ now is shifted downwards with the
increase of $t_{23}$, so that the energy region exhibiting the
unconventional Hall conductivity is enlarged. Therefore, no quantum
phase transition is observed in the energy region considered.

\begin{figure}
  \includegraphics[scale=0.38]{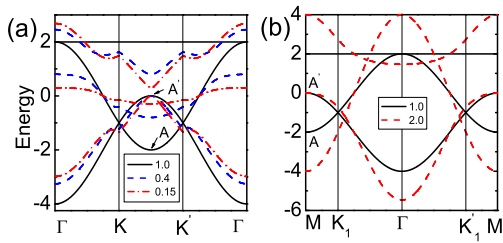}\\
  \caption{(color online) The energy bands for different $t_{23}$.
  (a) The energy bands for weak $t_{23}$ regime. Two typical energy bands are plotted along $\Gamma-K-K'-\Gamma$
  direction. The blue dash, and red dash-dot lines are corresponding to $t_{23}=0.4$ and $t_{23}=0.15$ respectively.
  (b) The energy bands for strong $t_{23}$ regime. The typical energy bands for $t_{23}=2$(the red dash lines)
  are plotted along $M-K_{1}-\Gamma-K_{2}-M'$ direction. In the two
  figures, the black solid lines are the energy bands for the isotropic case($t_{23}=1$).}
  \label{fig4}
\end{figure}

Finally, let us give a few comments on the possible experimental
realization of the theoretical prediction elaborated here. The
anisotropy of the hopping integral can be realized by the distortion
of the lattice due to the monoclinic distortion, such as in
$\mathrm{Cu_(3)V_{2}O_{7}(OH)_{2}\cdot2H_{2}O}$~\cite{Bert}, or by
the difference in orbital characters on the atomic sites in each
unit cell due to the Jahn-Teller effect, such as in
$\mathrm{Cs_{2}Cu_{3}CeF_{12}}$~\cite{Amemiya}. On the other hand,
the Kagom$\acute{e}$ lattice has been proposed to be realized by
implementing an optical lattice for ultra-cold
atoms~\cite{Santos,Ruostekoski}. In this regard, the ability to
conveniently control the physical parameters in the system
facilitates the realization of the anisotropy. It also worthwhile
to point out that the 2D lattice in the limit $t_{23}=0$ as shown in
Fig.3(c) is different from the Kagom$\acute{e}$ lattice, and may
provide an interesting model system for experimental investigation
on the special quantum dynamics demonstrated above.

In summary, we have study the quantum Hall effect on the anisotropic
Kagom$\acute{e}$ lattice. The anisotropy is introduced by assuming
one of the hopping integrals $t_{23}$ taking a different value. In
the weak $t_{23}$ ($t_{23}<t_{12}$) regime, we find a new type of
QHE characterized by the quantization rules for Hall conductivity
$\sigma_{xy}=2ne^{2}/h$ and Landau Levels $E(n)=\pm
v_{F}\sqrt{(n+1/2)\hbar Be}$, which is different from the known
types. This phase evolves from the unconventional QHE with
$\sigma_{xy}=4(n+1/2)e^{2}/h$ via a quantum phase transition, which
occurs successively from low to high energies with the decrease of
the hopping integral. This quantum phase transition is absent in the
strong $t_{23}$ regime. Possible experimental realization of the
theoretical prediction is also discussed.

\begin{acknowledgments}
We acknowledge valuable discussions with D.H.Lee and X.G.Wen. This
work was supported by the National Natural Science Foundation of
China (10525415 and 10874066), the Ministry of Science and Technology of
China (973 project Grants Nos.2006CB601002,2006CB921800, 2007CB925104,
and 2009CB929504).
\end{acknowledgments}


\begin{thebibliography}{99}
\bibitem{Prange} R. E. Prange, S. M. Girvin, The Quantum Hall Effect(Springer, New
York, 1990).
\bibitem{Novoselov1} K. S. Novoselov, A. K. Geim, S. V. Morozov, D. Jiang, M. I. Katsnelson, I. V. Grigorieva,
S. V. Dubonos, and A. A. Firsov, Nature(London) \textbf{438}, 197
(2005).
\bibitem{Zhang} Y. Zhang, Y. -W. Tan, H. L. Stormer, and P. Kim, Nature(London) \textbf{438}, 201 (2005).
\bibitem{Haldane} F. D. M. Haldane, Phys. Rev. Lett. \textbf{61}, 2015 (1988).
\bibitem{Zheng} Y. Zheng, and T. Ando, Phys. Rev. B \textbf{65}, 245420 (2002).
\bibitem{Gusynin} V. P. Gusynin, and S. G. Sharapov, Phys. Rev. Lett. \textbf{95}, 146801 (1999).
\bibitem{Sheng} D. N. Sheng, L. Sheng, and Z. Y. Weng, Phys. Rev. B
\textbf{73}, 233406 (2006).
\bibitem{Neto} A. H. Castro Neto, F. Guinea, N. M. R. Peres, K. S. Novoselov and A. K.
Geim, Rev. Mod. Phys. \textbf{81}, 109 (2009).
\bibitem{Novoselov2} K. S. Novoselov, E. Mccann, S. V. Morozov, V. I. Fal¡¯ko, M. I. Katsnelson, U. Zeitler,
D. Jiang, F. Schedin, and A. K. Geim, Nat. Phys. \textbf{2}, 177
(2006).
\bibitem{McCann} E. McCann and V. I. Fal¡¯ko, Phys. Rev. Lett. \textbf{96}, 086805 (2006).
\bibitem{Mielke} A. Mielke, J. Phys. A \textbf{24}, L73 (1991).
%\bibitem{Nagaosa1} K. Ohgushi, S. Murakami, and N. Nagaosa, Phys. Rev. B
%\textbf{62}, R6065 (2000).
%\bibitem{Nagaosa2} N. Nagaosa, Mater. Sci. Eng. B \textbf{84}, 58 (2001).
\bibitem{Sato} M. Sato, D. Tobe, and M. Kohmoto, Phys. Rev. B \textbf{78}, 235322 (2008).
\bibitem{Hasegawa} Y. Hasegawa, R. Konno, H. Nakano, and M. Kohmoto,
Phys. Rev. B \textbf{74}, 033413 (2006).
\bibitem{Dietl} P. Dietl, F. Pi$\acute{e}$chon, and G. Montambaux, Phys. Rev. Lett. \textbf{100}, 236405 (2008).
\bibitem{Kohmoto} M. Kohmoto and Y. Hasegawa, Phys. Rev. B \textbf{76}, 205402 (2007).
%\bibitem{Imai} Y. Imai and N. Kawakami, Phys. Rev. B \textbf{68}, 195103 (2003).
\bibitem{Mikitik} G. P. Mikitik and Yu. V. Sharlai, Phys. Rev. Lett. \textbf{82}, 2147 (1999).
%\bibitem{Bergman} D. L. Bergman, C. Wu, and L. Balents, Phys. Rev. B \textbf{78}, 125104 (2008).
\bibitem{Bert} F. Bert, D. Bono, P. Mendels, F. Ladieu, F. Duc, J.-C. Trombe, and P. Millet,
Phys. Rev. Lett. \textbf{95}, 087203 (2005).
\bibitem{Amemiya} T. Amemiya, M. Yano, K. Morita, I. Umegaki, T. Ono, H. Tanaka, K. Fujii, and H.
Uekusa, arXiv: 0906.1628 (2009).
\bibitem{Santos} L. Santos, M. A. Baranov, J. I. Cirac, H.-U. Everts, H. Fehrmann, and M.
Lewenstein, Phys. Rev. Lett. \textbf{93}, 030601 (1999).
\bibitem{Ruostekoski} J. Ruostekoski, arXiv: 0906.3042 (2009).
\end{thebibliography}
\end{document}